\documentclass{PoS}

\title{Light nuclei and nucleon form factors in $N_f=2+1$ lattice QCD}

\ShortTitle{Light nuclei and nucleon form factors in $N_f=2+1$ lattice QCD}

\author{\speaker{Takeshi Yamazaki}%
\ for PACS Collaboration\\
Faculty of Pure and Applied Sciences,
University of Tsukuba, Tsukuba, Ibaraki 305-8571, Japan\\
Center for Computational Sciences, University of Tsukuba,
Tsukuba, Ibaraki 305-8577, Japan\\
RIKEN Advanced Institute for Computational Science,
Kobe, Hyogo 650-0047, Japan\\
        E-mail: \email{yamazaki@het.ph.tsukuba.ac.jp}}

\abstract{We present our result of binding energy of light nuclei
with the nuclear mass number less than or equal to four
at the pion mass $m_\pi = 0.3$ GeV.
The simulations are performed in $N_f=2+1$ QCD with Iwasaki gauge
and a non-perturbative improved Wilson quark actions at the 
lattice spacing of $a = 0.09$ fm. 
We discuss the $m_\pi$ dependence of the binding energies
by comparing with previous results.
Furthermore, we show preliminary results for the axial charge
and the Dirac radius obtained from the nucleon form factors
at almost physical $m_\pi$.
}

\FullConference{The 33rd International Symposium on Lattice Field Theory\\
		 14 -18 July  2015\\
		 Kobe International Conference Center, Kobe, Japan}

\begin{document}

\section{Introduction}

The formation of the helium nuclei was examined 
in quenched lattice QCD at the pion mass
$m_\pi = 0.8$ GeV in Ref.~\cite{Yamazaki:2009ua}.
After this calculation, several calculations~\cite{Yamazaki:2011nd,Beane:2011iw,Beane:2012vq,Yamazaki:2012hi} 
of nucleus with the nuclear mass number less than or equal to four have been
reported.
However, there are discrepancies between the lattice calculations and the experiments:
the binding energy of the nucleus is larger than the experimental one,
and the dineutron, which does not exist in nature, is formed in
the lattice calculation.

One possible explanation of the differences from the experiment is 
systematic errors coming from larger $m_\pi$ than the physical one
in the calculations.
To check this possibility, we calculate the nuclei 
in the $^4$He, $^3$He, two-nucleon spin-triplet($^3$S$_1$) and 
spin-singlet($^1$S$_0$) channels in
$N_f = 2+1$ QCD at $m_\pi = 0.3$ GeV.
The result of this calculation has already been published 
in Ref.~\cite{Yamazaki:2015asa}.

Furthermore, we calculate the nucleon form factors at almost physical $m_\pi$
on a large volume of the spatial extent of 8 fm.
Preliminary results of some physical quantities obtained from
the nucleon form factors are also presented in this report.

\section{Light nuclei}

For the nuclei calculation at $m_\pi = 0.3$ GeV, we generate
the gauge configurations with the Iwasaki gauge action and
a non-perturbative improved Wilson quark action in $N_f = 2 + 1$ QCD
at the lattice spacing of $a=0.09$ fm
on the two lattice sizes, $L^3\times T = 48^3\times 48$ and $64^3 \times 64$,
corresponding to the spatial extent of 4.3 and 5.8 fm, respectively.
The correlation functions in
the $^4$He, $^3$He, $^3$S$_1$ and $^1$S$_0$ channels
are measured on the gauge configuration.
In the measurement an exponential smeared quark source and 
point sink operators are employed in all the channels.  
The details of the simulation are explained in 
the published paper~\cite{Yamazaki:2015asa}.

We identify bound state in each channel
from the volume dependence of the energy 
shift~\cite{Luscher:1990ux,Beane:2003da,Sasaki:2006jn},
$\Delta E_L = E_{N_N} - N_N m_N$,
where $E_{N_N}$ is the ground state energy of $N_N$-nucleon channel
and $m_N$ is the nucleon mass on a volume of the spatial extent of $L$.
We observe that $\Delta E_L$ is nonzero in the infinite volume limit
for all the channels.
Thus, we conclude that there is nucleus in each channel,
and determine the binding energy from
the value of $\Delta E_L$ in the infinite volume limit.

\subsection{Helium channels}

\begin{figure}[!t]
\begin{tabular}{cc}
\includegraphics*[angle=0,width=0.49\textwidth]{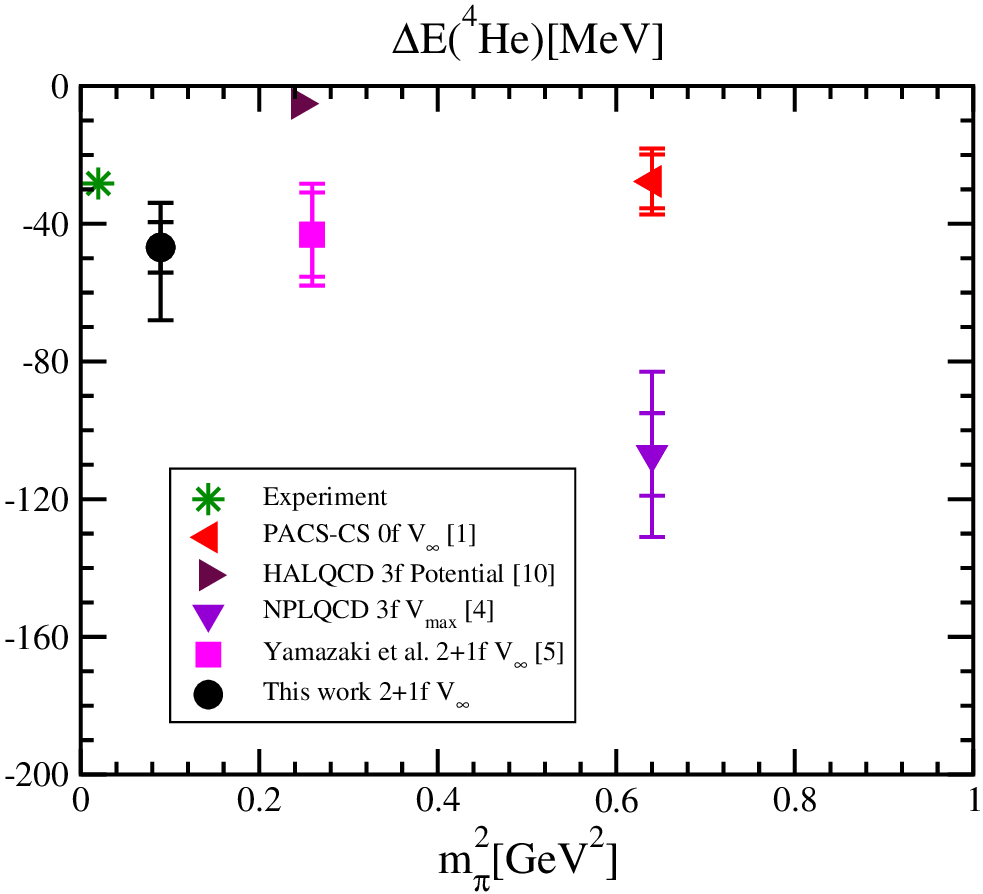}
&
\includegraphics*[angle=0,width=0.49\textwidth]{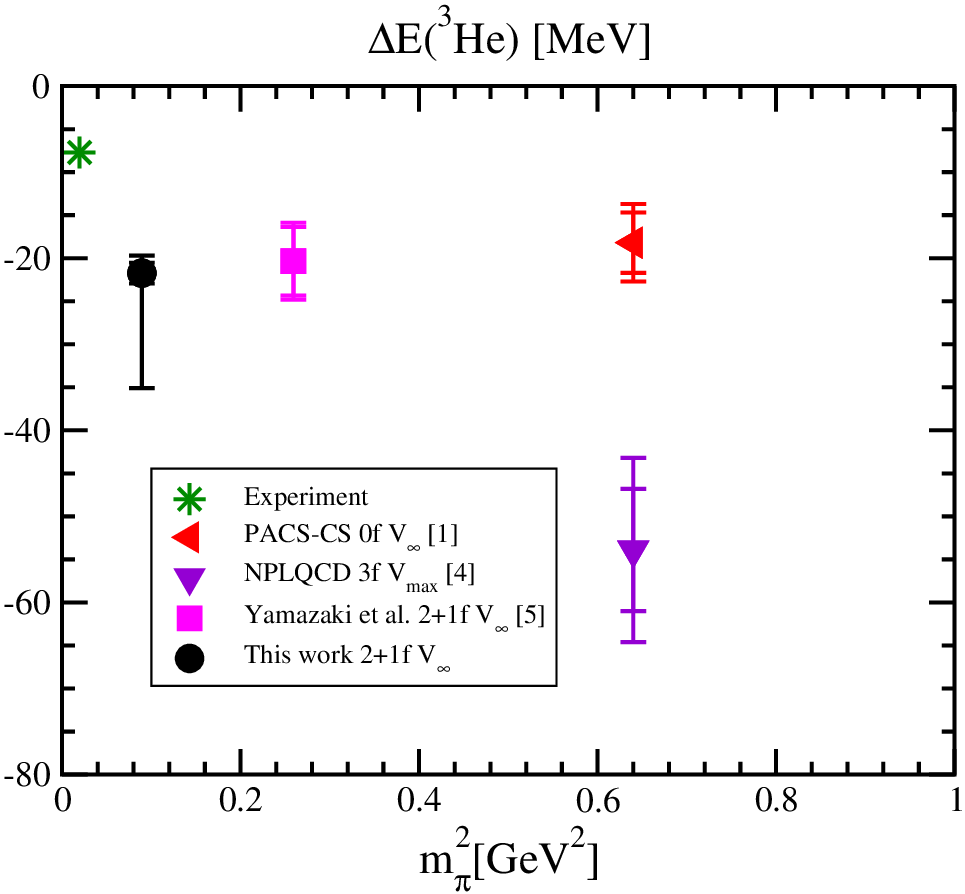}
\vspace*{-4mm}
\end{tabular}
\caption{
Binding energies for $^4$He (left) and $^3$He (right) channels as a function
of $m_\pi^2$.
The inner bar of each data denotes the statistical error and 
the outer bar represents the 
total error with the statistical and 
systematic ones added in quadrature.
Filled circle is the result of this calculation.
Left, right, down triangles, and square symbols are 
the previous results in
Refs.~\cite{Yamazaki:2009ua,Inoue:2011ai,Beane:2012vq,Yamazaki:2012hi}, 
respectively.
\label{fig:He}
}
\end{figure}

The results of the binding energy at $m_\pi = 0.3$ GeV
for the $^4$He and $^3$He channels are shown in each panel of 
Fig.~\ref{fig:He} together with the previous results and the 
experimental values.
In the $^4$He channel 
the binding energy of $m_\pi=0.3$ GeV
is similar in magnitude with our previous results for
$N_f = 2+1$ at $m_\pi = 0.51$ GeV~\cite{Yamazaki:2012hi} and 
$N_f = 0$ at $m_\pi = 0.80$ GeV~\cite{Yamazaki:2011nd}.
Furthermore, the result is consistent with the experiment
within 1.5 $\sigma$ using the upper total error, where
the statistical and systematic errors are added in quadrature.

The result of $m_\pi=0.3$ GeV in the $^3$He channel has 
the large systematic error, since
the effective $\Delta E_L$ does not have
clear plateau in both the volumes.
Our result and the ones in the previous calculations have larger binding energy
than the experimental data, which might be caused by the larger 
$m_\pi$ in the calculations than the physical one.

\subsection{Two-nucleon channels}

\begin{figure}[!t]
\begin{tabular}{cc}
\includegraphics*[angle=0,width=0.49\textwidth]{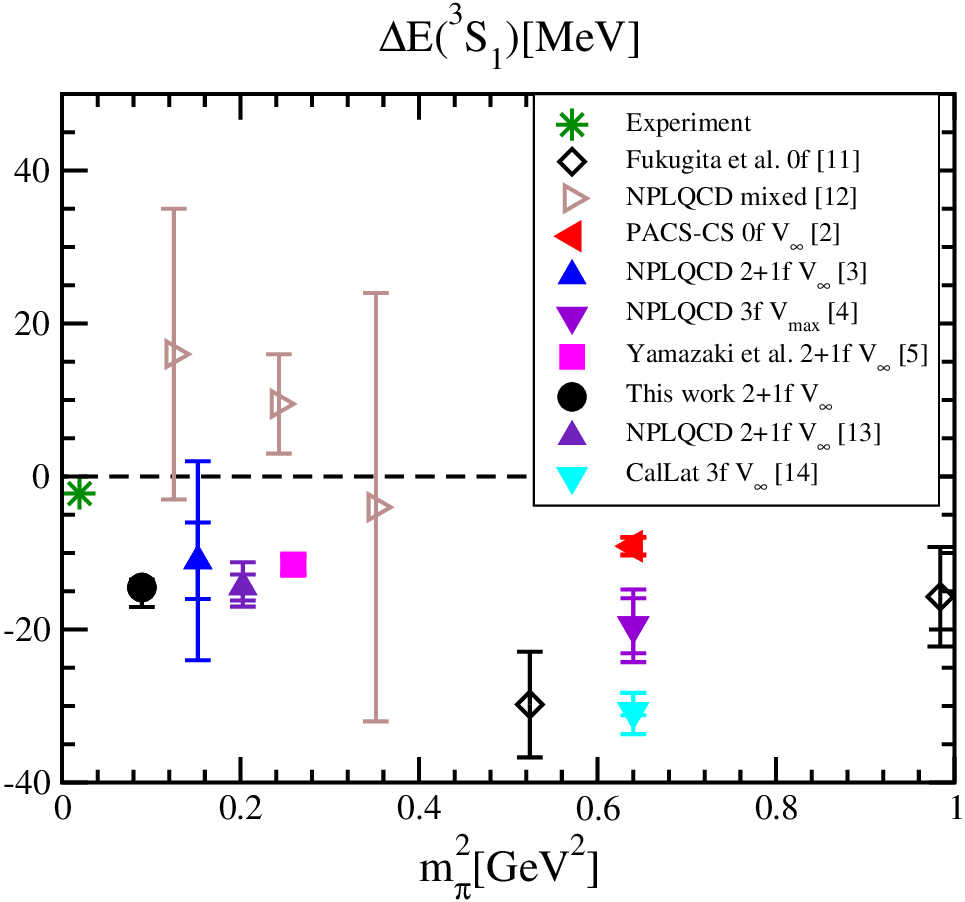}
&
\includegraphics*[angle=0,width=0.49\textwidth]{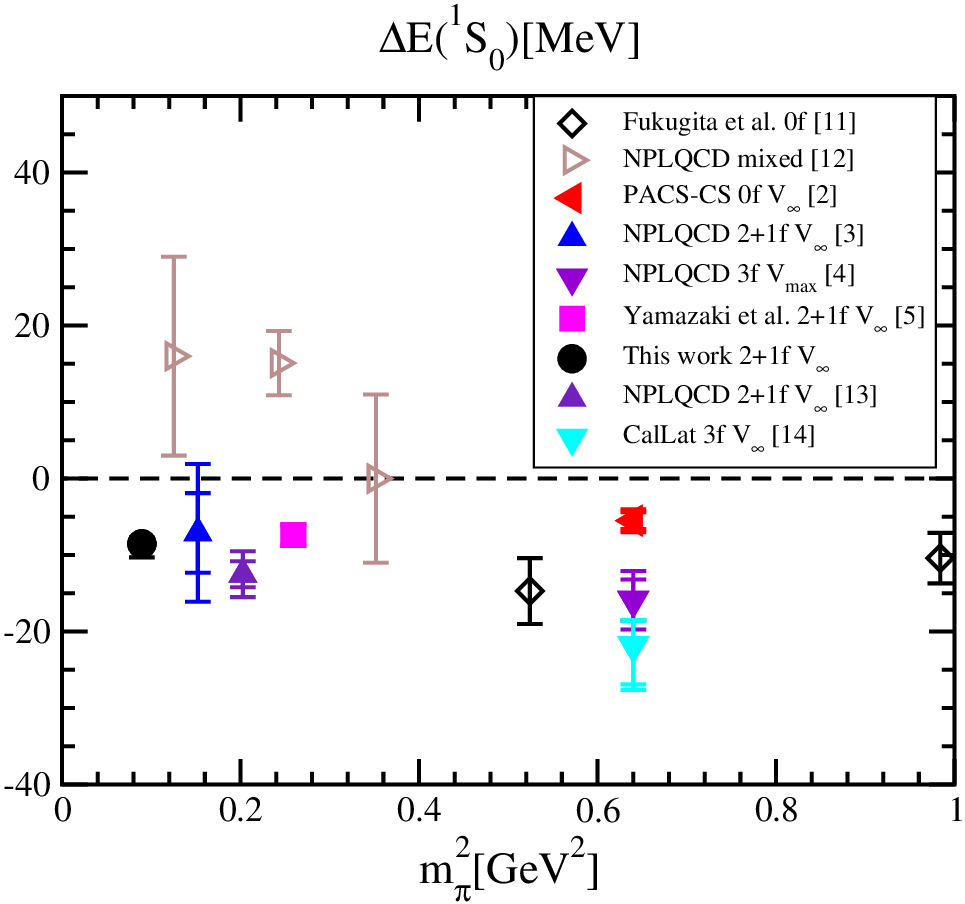}
\vspace*{-4mm}
\end{tabular}
\caption{
Binding energies for two-nucleon $^3$S$_1$ (left) and $^1$S$_0$ (right) 
channels as a function of $m_\pi^2$.
The inner bar of each data denotes the statistical error and 
the outer bar represents the 
total error with the statistical and 
systematic ones added in quadrature.
Filled circle is the result of this calculation.
Diamond, left, right, up, down triangles, and square symbols are 
the previous results in 
Refs.~\cite{Fukugita:1994ve,Beane:2006mx,Yamazaki:2011nd,Beane:2011iw,Beane:2012vq,Yamazaki:2012hi}, respectively.
Violet up and cyan down triangles are recent results by
NPLQCD~\cite{Orginos:2015aya} and CalLat~\cite{Berkowitz:2015eaa} 
Collaborations.
\label{fig:NN}
}
\end{figure}

The results of the $^3$S$_1$ and $^1$S$_0$ channels in the two-nucleon
systems are summarized in Fig.~\ref{fig:NN}.
The circle symbol in each panel is the result
of $m_\pi = 0.3$ GeV.
The open symbols express the data calculated on one volume,
so that it is not clear whether the calculated state is bound or attractive
scattering state.
All the closed symbols are confirmed that 
the existence of nucleus by the investigation
of volume dependence of $\Delta E_L$.
The result of this work in both the channels is reasonably consistent with
our previous result at $m_\pi = 0.5$ GeV and also 
the result at $m_\pi = 0.45$ GeV of NPLQCD Collaboration.

For comparison with the experiment, the lattice results in the 
$^3$S$_1$ channel has
a factor five to ten times larger binding energy than the experiment.
Furthermore, all the lattice results in the $^1$S$_0$ channel
has the bound state, which is not observed in the experiment.
It might be caused by the larger $m_\pi$ than the physical one
in the calculations,
while we have not observed the expected $m_\pi$ dependences
such that the binding energy 
approaches to the experimental value in the $^3$S$_1$ channel,
and it decreases toward the chiral limit in the $^1$S$_0$ channel.
It is an important future work to check the consistency with the experiment
in the physical $m_\pi$.
For this purpose 
we start calculation at almost physical point of $m_\pi = 0.145$ GeV
in the two-nucleon channels as well as the helium channels.

\section{Nucleon form factors}

Nucleon form factors are measured on the $N_f = 2+1$ QCD 
gauge configurations generated at almost physical point of 
$m_\pi = 0.145$ GeV on
a large volume of the spatial extent of 8 fm.
The lattice spacing is about $0.086$ fm.
For the configuration generation and the form factor measurement,
we employ Iwasaki gauge action and
the stout smeared $O(a)$ improved Wilson quark action.
The results of hadron spectrum with this gauge configuration are presented in
this conference~\cite{Ukita:2015xx}.

The nucleon form factors are obtained from the nucleon
three-point functions of the local vector and axial vector currents.
We adopt the sequential source method to calculate the 
three-point functions using an exponential smeared quark fields
in the nucleon operators in the source and sink time slices,
which are separated by 15 lattice units.
The three-point functions are calculated from only connected diagrams,
so that we calculate the isovector part of the nucleon form factors.
We use 104 configurations, and perform 64 measurements
of the three-point functions per configuration.
All the results presented in the next subsections are preliminary.

\subsection{$Z_V$ and Axial charge}

\begin{figure}[!t]
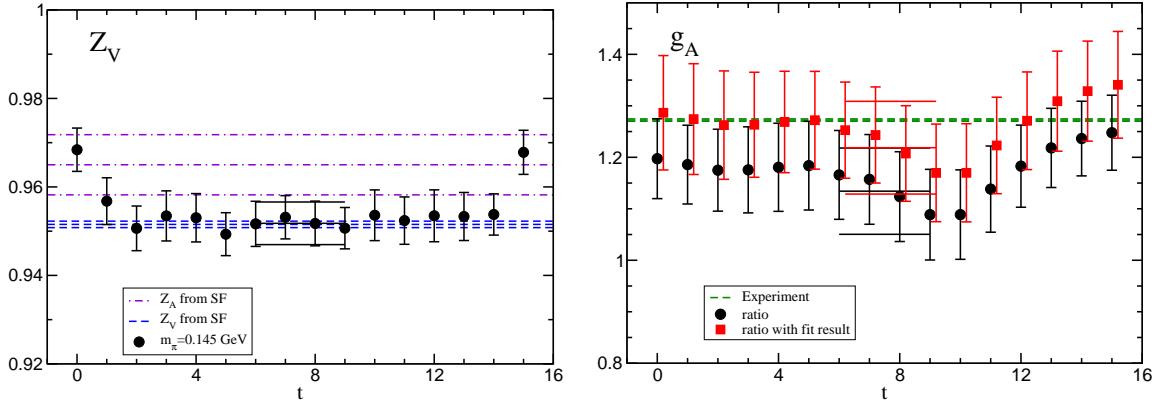

\begin{tabular}{cc}
\includegraphics*[angle=0,width=0.49\textwidth]{Figs/zv.eps}
&
\includegraphics*[angle=0,width=0.49\textwidth]{Figs/ga.eps}
\vspace*{-4mm}
\end{tabular}
\caption{
Preliminary results of renormalization factor of vector current $Z_V$(left)
and axial charge $g_A$(right) at $m_\pi = 0.145$ GeV.
The solid lines express constant fit results with error band.
In the left panel, $Z_V$(dashed line) and $Z_A$(dot-dashed line) in
Schr\"odinger functional scheme~\cite{Ishikawa:2015xx} with the 
error band are also plotted.
In the right panel, the experimental value is expressed by dashed line,
and the two results are explained in text.
\label{fig:zv_ga}
}
\end{figure}

Using the vector three-point function at the zero momentum transfer,
we evaluate the renormalization factor of the vector current $Z_V$
shown in the left panel of Fig.~\ref{fig:zv_ga}.
The result is consistent with $Z_V$ and the renormalization factor
of the axial vector current $Z_A$ calculated in Schr\"odinger functional
scheme within 2\%. 
This means that the chiral symmetry breaking effect is small in this
quantity.
The calculation in Schr\"odinger functional
scheme is presented in this conference~\cite{Ishikawa:2015xx}.

The axial charge $g_A$ is evaluated from the three-point function
of the axial vector current at zero momentum transfer presented in
the right panel of Fig.~\ref{fig:zv_ga}.
We use $Z_A$ shown in the left panel of Fig.~\ref{fig:zv_ga}
to renormalize the bare $g_A$.
The bare $g_A$ is calculated from the ratio of the three-
and two-point functions, and also the same ratio but with
the two-point function reconstructed from its fit result.
The discrepancy of the two results shown in the figure is
a systematic error of this calculation.
While the statistical errors are still large, the results are comparable
to the experimental result.

\subsection{Vector form factors}

\begin{figure}[!t]
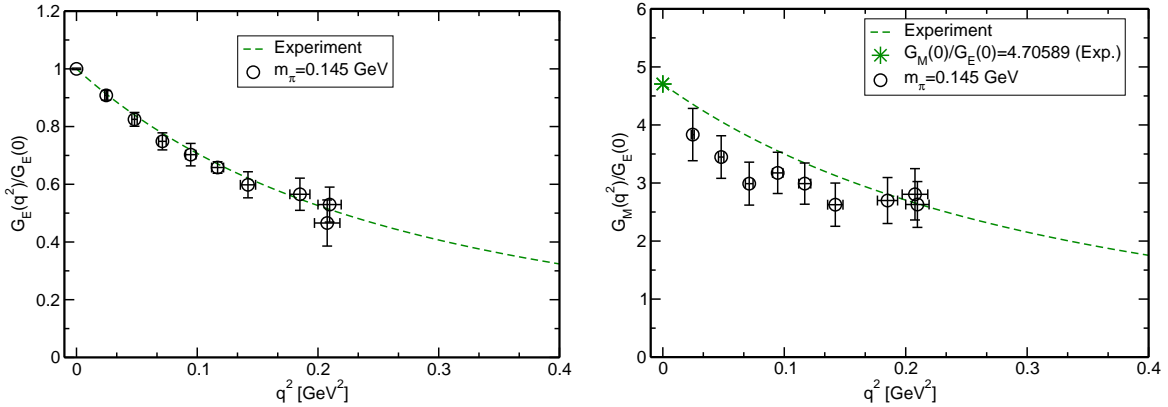

\begin{tabular}{cc}
\includegraphics*[angle=0,width=0.5\textwidth]{Figs/Ge_mom9_104_abs.eps}
&
\includegraphics*[angle=0,width=0.49\textwidth]{Figs/Gm_mom9_104_abs.eps}
\vspace*{-4mm}
\end{tabular}
\caption{
Preliminary results of electric and magnetic Sachs form factors,
$G_E(q^2)$(left) and $G_M(q^2)$(right) at $m_\pi = 0.145$ GeV.
The dashed lines and star symbol express experimental results.
\label{fig:Ge_Gm}
}
\end{figure}

The preliminary result of the electric and magnetic Sachs form factors,
$G_E(q^2)$ and $G_M(q^2)$, is
presented in the left and right panels of Fig.~\ref{fig:Ge_Gm},
respectively.
We also plot the experimental curves in the figure.
The form factors are evaluated from the three-point function
of the vector current at the nonzero momentum transfer $q^2$
in the same way as in Ref.~\cite{Yamazaki:2009zq}.
While statistical errors are large, it is encouraging that
$G_E(q^2)$ agrees with the experiment, and 
$G_M(q^2)$ is roughly consistent with the experimental curve.

\begin{figure}[!t]
\hfil
\includegraphics*[angle=0,width=0.45\textwidth]{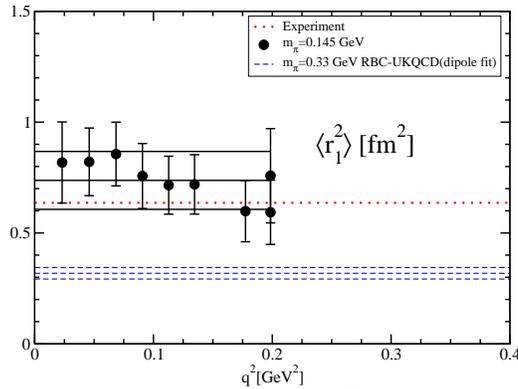}
\vspace*{-4mm}
\caption{
Preliminary results of mean square Dirac radius $\langle r_1^2 \rangle$
at $m_\pi = 0.145$ GeV.
The circle symbols are effective $\langle r_1^2 \rangle$,
and the solid lines are the dipole fit of $F_1(q^2)$ 
with the error band.
The dotted and dashed lines represent experimental result and
the previous result at $m_\pi = 0.33$ GeV~\cite{Yamazaki:2009zq}, 
respectively.
\label{fig:r_1}
}
\end{figure}

The Dirac and Pauli form factors, $F_1(q^2)$ and $F_2(q^2)$,
are evaluated from $G_E(q^2)$ and $G_M(q^2)$ by solving the linear equations,
\begin{equation}
G_E(q^2)=F_1(q^2) - \frac{q^2}{4m_N^2}F_2(q^2),\ \ \ 
G_M(q^2) = F_1(q^2) + F_2(q^2).
\end{equation}
The mean square Dirac radius $\langle r_1^2 \rangle$
is estimated from our result of $F_1(q^2)$ assuming the dipole
form, $F_1(q^2) = 1/(1+\langle r_1^2 \rangle q^2/12)^2$.
At each $q^2$ the effective $\langle r_1^2 \rangle$~\cite{Sasaki:2007gw} 
is evaluated from the dipole form,
\begin{equation}
\langle r_1^2 \rangle = 
\frac{12}{q^2}\left(\sqrt{\frac{1}{F_1(q^2)}}-1\right).
\end{equation}
The result of the effective $\langle r_1^2 \rangle$ is plotted 
in Fig.~\ref{fig:r_1}.
Since the typical result of $\langle r_1^2 \rangle$
at $m_\pi > 0.3$ GeV is smaller
than the experiment, for example the one at $m_\pi=0.33$ GeV
in Ref.~\cite{Yamazaki:2009zq} plotted in the figure,
it is encouraging that our result is 
consistent with the experiment within the large statistical error.
The $q^2$ dependence of the effective $\langle r_1^2 \rangle$ 
is reasonably flat.
Thus, we obtain a consistent result from a dipole fit of $F_1(q^2)$
plotted in Fig.~\ref{fig:r_1}.

\section{Summary}

We have calculated the binding energies of nuclei in the $^4$He, $^3$He,
two-nucleon $^3$S$_1$, and $^1$S$_0$ channels at $m_\pi = 0.3$ GeV
in $N_f = 2+1$ lattice QCD.
We have obtained consistent results with the ones in our calculation
at $m_\pi = 0.5$ GeV in all the channels,
and the ones in the recent calculation by NPLQCD Collaboration at $m_\pi =0.45$
GeV in the two-nucleon channels.
We, however, have not observed $m_\pi$ dependences of the binding energy
approaching to the experimental values, especially in the two-nucleon channels.
It is an important future work 
to check the consistency with the experimental result 
from calculations in the physical $m_\pi$.
For this direction, we have already started the calculation at 
$m_\pi = 0.145$ GeV in all the channels.
Other possible sources of systematic error are 
finite lattice spacing effect and excited state contaminations
discussed in Ref.~\cite{Yamazaki:2015asa}.

We also have calculated the nucleon form factors at almost physical
$m_\pi$.
While the results are still preliminary, the encouraging results are obtained in
$g_A$, $G_E(q^2)$, $G_M(q^2)$, and $\langle r_1^2 \rangle$,
which are roughly consistent with the experiment within
the large statistical error.
An important future work is to reduce statistical error, and
compare with the experiment and
also the recent lattice calculations near physical $m_\pi$,
which are summarized in 
the recent plenary talks~\cite{Constantinou:2014tga,Zanotti:2015xx}.

\section*{Acknowledgements}
Numerical calculations for the present work have been carried out
on the FX10 cluster system at Information Technology Center
of the University of Tokyo, 
on the T2K-Tsukuba cluster system and HA-PACS system and COMA system
at University of Tsukuba,
on the K computer at RIKEN Advanced Institute for Computational Science,
on HOKUSAI GreatWave at Advanced Center for Computing 
and Communication of RIKEN,
and on the computer facilities of the Research Institute for Information 
Technology of Kyushu University. 
This research used computational resources of the HPCI system provided 
by Information Technology Center
of the University of Tokyo, 
Institute for Information Management and Communication 
of Kyoto University,
the Information Technology Center of Nagoya University,
and RIKEN Advanced Institute for Computational Science
through the HPCI System Research Project 
(Project ID: hp120281, hp130023, hp140209, hp140155, hp150135).
We thank the colleagues in the PACS Collaboration for helpful
discussions and providing us the code used in this work.
This work is supported in part by MEXT SPIRE Field 5 and JICFuS,
and also by Grants-in-Aid for Scientific Research
from the Ministry of Education, Culture, Sports, Science and Technology 
(Nos. 22244018, 25800138) and 
Grants-in-Aid of the Japanese Ministry for Scientific Research on Innovative 
Areas (Nos. 20105002, 21105501, 23105708).

\bibliography{lat15}

\providecommand{\href}[2]{#2}\begingroup\raggedright\begin{thebibliography}{10}

\bibitem{Yamazaki:2009ua}
{\bf PACS-CS Collaboration}, T.~Yamazaki, Y.~Kuramashi, and A.~Ukawa, {\em
  Phys.Rev.} {\bf D81} (2010) 111504.

\bibitem{Yamazaki:2011nd}
{\bf PACS-CS Collaboration}, T.~Yamazaki, Y.~Kuramashi, and A.~Ukawa, {\em
  Phys. Rev.} {\bf D84} (2011) 054506.

\bibitem{Beane:2011iw}
{\bf NPLQCD Collaboration}, S.~Beane et~al., {\em Phys. Rev.} {\bf D85} (2012)
  054511.

\bibitem{Beane:2012vq}
{\bf NPLQCD Collaboration}, S.~Beane, E.~Chang, S.~Cohen, W.~Detmold, H.~Lin,
  et~al., {\em Phys.Rev.} {\bf D87} (2013), no.~3 034506.

\bibitem{Yamazaki:2012hi}
T.~Yamazaki, K.-i. Ishikawa, Y.~Kuramashi, and A.~Ukawa, {\em Phys.Rev.} {\bf
  D86} (2012) 074514.

\bibitem{Yamazaki:2015asa}
T.~Yamazaki, K.-i. Ishikawa, Y.~Kuramashi, and A.~Ukawa, {\em Phys. Rev.} {\bf
  D92} (2015), no.~1 014501.

\bibitem{Luscher:1990ux}
M.~L{\"u}scher, {\em Nucl. Phys.} {\bf B354} (1991) 531--578.

\bibitem{Beane:2003da}
S.~R. Beane, P.~F. Bedaque, A.~Parreno, and M.~J. Savage, {\em Phys. Lett.}
  {\bf B585} (2004) 106--114.

\bibitem{Sasaki:2006jn}
S.~Sasaki and T.~Yamazaki, {\em Phys. Rev.} {\bf D74} (2006) 114507.

\bibitem{Inoue:2011ai}
{\bf HALQCD Collaboration}, T.~Inoue et~al., {\em Nucl. Phys.} {\bf A881}
  (2012) 28--43.

\bibitem{Fukugita:1994ve}
M.~Fukugita, Y.~Kuramashi, M.~Okawa, H.~Mino, and A.~Ukawa, {\em Phys. Rev.}
  {\bf D52} (1995) 3003--3023.

\bibitem{Beane:2006mx}
S.~R. Beane, P.~F. Bedaque, K.~Orginos, and M.~J. Savage, {\em Phys. Rev.
  Lett.} {\bf 97} (2006) 012001.

\bibitem{Orginos:2015aya}
{\bf NPLQCD Collaboration}, K.~Orginos, A.~Parreno, M.~J. Savage, S.~R. Beane,
  E.~Chang, and W.~Detmold, \href{http://arxiv.org/abs/1508.07583}{{\tt
  arXiv:1508.07583}}.

\bibitem{Berkowitz:2015eaa}
E.~Berkowitz, T.~Kurth, A.~Nicholson, B.~Joo, E.~Rinaldi, M.~Strother, P.~M.
  Vranas, and A.~Walker-Loud, \href{http://arxiv.org/abs/1508.00886}{{\tt
  arXiv:1508.00886}}.

\bibitem{Ukita:2015xx}
{\bf PACS Collaboration}, N.~Ukita et~al., {\em PoS} {\bf LATTICE2015} (2015)
  075.

\bibitem{Ishikawa:2015xx}
{\bf PACS Collaboration}, K.-i. Ishikawa et~al., {\em PoS} {\bf LATTICE2015}
  (2015) 271.

\bibitem{Yamazaki:2009zq}
T.~Yamazaki, Y.~Aoki, T.~Blum, H.-W. Lin, S.~Ohta, S.~Sasaki, R.~Tweedie, and
  J.~Zanotti, {\em Phys. Rev.} {\bf D79} (2009) 114505.

\bibitem{Sasaki:2007gw}
S.~Sasaki and T.~Yamazaki, {\em Phys. Rev.} {\bf D78} (2008) 014510.

\bibitem{Constantinou:2014tga}
M.~Constantinou, {\em PoS} {\bf LATTICE2014} (2015) 001.

\bibitem{Zanotti:2015xx}
J.~Zanotti, {\em PoS} {\bf LATTICE2015} (2015) 005.

\end{thebibliography}\endgroup

\end{document}